\documentclass[dvipdfmx,aps,twocolumn,showpacs]{revtex4}
\usepackage{amsmath,amssymb,graphicx}

\usepackage[usenames]{color}

\newcommand{\Ne}{N_\mathrm{e}}
\newcommand{\Smax}{S_\mathrm{max}}
\newcommand{\Chsr}{C_\mathrm{hs}(r)}
\newcommand{\tChsr}{\tilde{C}_\mathrm{hs}(r)}
\newcommand{\nhi}{n_\mathrm{h}(i)}

\begin{document}

\title{Ferromagnetic clouds caused by hole motion in a one-dimensional $t$-$J$ model}

\author{Kazuhiro Sano}
\affiliation{Department of Physics Engineering, Mie University, Tsu, Mie 514-8507, Japan} 
\author{Ken'ichi Takano}
\affiliation{Toyota Technological Institute, Tenpaku-ku, Nagoya 468-8511, Japan}

\date{\today} %date

\begin{abstract}
The one-dimensional $t_1$-$t_2$-$J_1$-$J_2$ model is examined in the one-hole case, in which the total number of electrons is one less than the number of the lattice sites. 
The ground-state phase diagram includes a series of partial ferromagnetic phases, which are stacked in a regime of positive and small $J_1$. 
We find that the ground state in each of these partial ferromagnetic phases includes a ferromagnetic cloud, which is a multiple-spin bound state together with the hole. 
The ferromagnetic cloud is a large magnetic polaron with a heavy mass in a single-band electronic system and is supposedly formed as a result of Nagaoka ferromagnetism which locally works around the hole. 
\end{abstract} 


\maketitle

\section{Introduction}\label{intro}

The motion of a hole injected into a Mott insulator may play a decisive role in the electronic states of various phenomena such as metal-insulator transitions, magnetism, and superconductivity. 
The discovery of cuprate superconductors has stimulated the research on electronic ground states induced by hole motion. 
In particular, such electronic ground states in low dimensional systems may be exotic, owing to strong quantum fluctuations.\cite{hole-ref} 
Despite such intriguing possibilities, we have not yet arrived at a sufficient understanding of the relevance of  hole motion to electronic states, since the problem is inherently a quantum many-body problem of strongly correlated electrons surrounding the hole. 
It is therefore worthwhile to thoroughly investigate the problem of hole motion using simplified models by numerical methods. 

The $t$-$J$ model is an important model that represents strongly correlated electrons in a nearly half-filled band. 
The one-dimensional (1D) $t_1$-$t_2$-$J_1$-$J_2$ model is a kind of the $t$-$J$ model, which includes the first and second transfer (exchange) parameters, $t_1$ and $t_2$ ($J_1$ and $J_2$), respectively. 
In previous works, we examined this model in the one-hole case, in which the total number of electrons is one less than the number of lattice sites.\cite{Doi1992,Sano2011} 
Through numerical diagonalization with up to 15 lattice sites, we calculated the ground-state phase diagram in the $J_1$-$t_2$ plane, with the condition that $J_2 = (t_2/t_1)^2 J_1$ and with units of $t_1 =1$. 
The phase diagram includes a large characteristic phase with total spin $S=1$ in the quadrant $t_2 < 0$ and for $J_1 > 0$. 
We have determined that the ground state includes a tightly bound state of the hole and a triplet electron pair in a singlet background. 
This bound state moves as a composite particle, or as a small magnetic polaron.\cite{mpolaron} 

To understand why this bound state is formed, we consider two electrons in a triangle lattice with just three sites, and with two $t_1$ and one $t_2$ transfers. 
The ground state of this system is a triplet state for $t_2 < 0$ and a singlet state for $t_2 > 0$. 
In a larger lattice, the same mechanism works locally near the hole to gain energy. 

Apart from the large phase of $S=1$, we have found a series of narrow partial ferromagnetic phases with $S=2, 3, 4, \cdots$ in the same phase diagram\cite{Sano2011}; 
Magnified versions of the phase diagrams have been created for larger systems and are shown in Fig. \ref{phase_fbc} of the present paper. 
These partial ferromagnetic phases are very narrow and accumulate on the lower side of $J_1$ near the $t_2$ axis. 
These phases may be a new type of partial ferromagnetic phase. 
However, the narrowness of the phases could not exclude the possibility  that they are  artifacts, arising because of the small system sizes used in the numerical calculations. 
It is therefore very important to determine whether these phases survive in the large system-size limit. 

In this paper, we examine the ground states of the narrow partial ferromagnetic phases with spin $S$ larger than 1 in the 1D $t_1$-$t_2$-$J_1$-$J_2$ model. 
We perform numerical diagonalization for various system sizes and various set of parameters. 
By extrapolating with respect to system size, we ensure that these phases are not artifacts of the finite-size systems used in the numerical calculations. 
We then analyze the numerically obtained ground states for $S = 2, 3, 4, \cdots$ through their correlation functions and wave functions. 
The analysis shows that the $2S$ spins form a loose multi-spin bound state together with the hole in the spin $S$ partial ferromagnetic phase. 
This multiple-spin bound state is a kind of large magnetic polaron in a single band electronic system and we later call it the ferromagnetic cloud. 

This paper is organized as follows. 
In Sec. \ref{ham}, we introduce and explain the Hamiltonian of the 1D $t_1$-$t_2$-$J_1$-$J_2$ model. 
In Sec. \ref{pfphase}, we present the ground-state phase diagram of the model and examine the system-size dependence of a series of partial ferromagnetic phases near the $t_2$ axis. 
In Sec. \ref{msbstate}, we numerically examine the ground states of the partial ferromagnetic phases to show that they include multiple-spin bound states with the hole. 
Section \ref{disc} is devoted to discussion.  

\bigskip\bigskip

%%%%%%%%%%%%%%%%%%%%%%
\section{Hamiltonian}\label{ham}

We consider electronic states that are formed by the motion of a hole injected into an antiferromagnetic spin chain. 
This situation is represented by the 1D $t_1$-$t_2$-$J_1$-$J_2$ model. 
The Hamiltonian is written as 
%-------------------------------------------------------------
\begin{align} %Eq_(1)
H &= -t_{1}\sum_{i,\sigma}(c_{i,\sigma}^{\dagger} c_{i+1,\sigma}+h.c.) 
\displaybreak[0] \nonumber\\
&-t_{2}\sum_{i,\sigma}(c_{i,\sigma}^{\dagger} c_{i+2,\sigma}+h.c.) \displaybreak[0] \nonumber\\
&+J_{1}\sum_{i} \left({\bf S}_{i} \cdot {\bf S}_{i+1}-\frac{1}{4}n_in_{i+1} \right)  
\displaybreak[0] \nonumber\\
&+J_{2}\sum_{i} \left({\bf S}_{i} \cdot {\bf S}_{i+2}-\frac{1}{4}n_{i}n_{i+2} \right) 
\label{tJ-Ham} 
\end{align} 
%-------------------------------------------------------------
with 
%-------------------------------------------------------------
\begin{align} %Eq_(2)(3)
\mathbf{S}_{i} &= \sum_{\sigma \sigma'} 
c_{i,\sigma}^{\dagger} ({\boldsymbol \tau})_{\sigma \sigma'} c_{i,\sigma'} , \\
n_{i} &= \sum_\sigma c_{i,\sigma}^{\dagger}c_{i,\sigma} , 
\end{align}
%-------------------------------------------------------------
where $c^{\dagger}_{i,\sigma}$ ($c_{i,\sigma}$) is the creation (annihilation) operator of an electron with spin $\sigma$ at site $i$ and ${\boldsymbol \tau} = (\tau_x, \tau_y, \tau_z)$ is of the Pauli matrices. 
The infinite on-site repulsion is taken into account by removing states with doubly-occupied sites from the Hilbert space. 

Hamiltonian (\ref{tJ-Ham}) includes four parameters, $t_1$, $t_2$, $J_1$, and $J_2$, with dimensions of energy. 
Parameters $t_1$ and $t_2$ are the transfer energies between nearest-neighbor sites and next-nearest-neighbor sites, respectively. 
Parameters $J_1$ and $J_2$ are the exchange energies between electrons at nearest-neighbor sites and at next-nearest-neighbor sites, respectively. 
We are interested in the case of $J_1 > 0$, where the electronic spins intend to form a singlet state. 
We also concentrate on the case of negative $t_2$, where thin partial ferromagnetic phases appear, as presented in a previous paper.\cite{Sano2011} 

We reduce the number of parameters in the numerical calculations by imposing the condition $J_2 = (t_2/t_1)^2 J_1$. 
This condition preserves the approximate correspondence of the present model, with positive, small $J_1$ and $J_2$, to the Hubbard model with large on-site repulsion. 
We also set $t_1=1$, which is the energy unit. 
The independent parameters are then $J_1$ and $t_2$. 

We denote the number of lattice sites as $N$ and the number of electrons as $\Ne$ $(= \sum_i n_i)$ below. 
The one-hole case is represented by $\Ne = N - 1$. 
We also denote the total electronic spin as $\mathbf{S} = \sum_i \mathbf{S}_{i}$. 
The eigenvalues of the $z$-component and the magnitude of $\mathbf{S}$ are represented by $S_z$ and $S$, respectively. 

%%%%%%%%%%%%%%%%%%%%%%%%
\section{Phase diagram --- Existence of partial ferromagnetic phases ---}\label{pfphase}

\begin{figure}[b]
\begin{center}
\includegraphics[width=0.7 \linewidth]{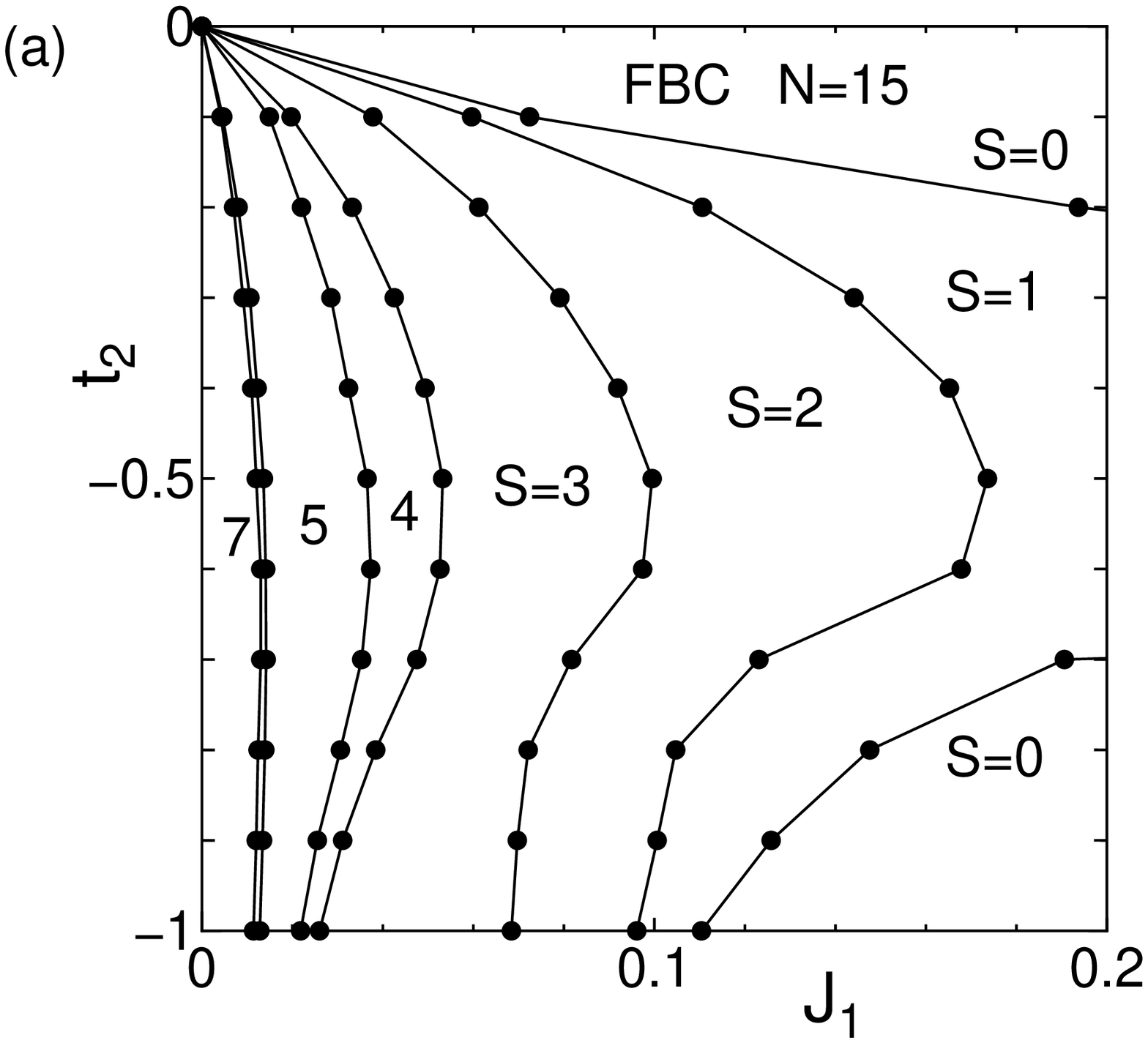}
\includegraphics[width=0.7 \linewidth]{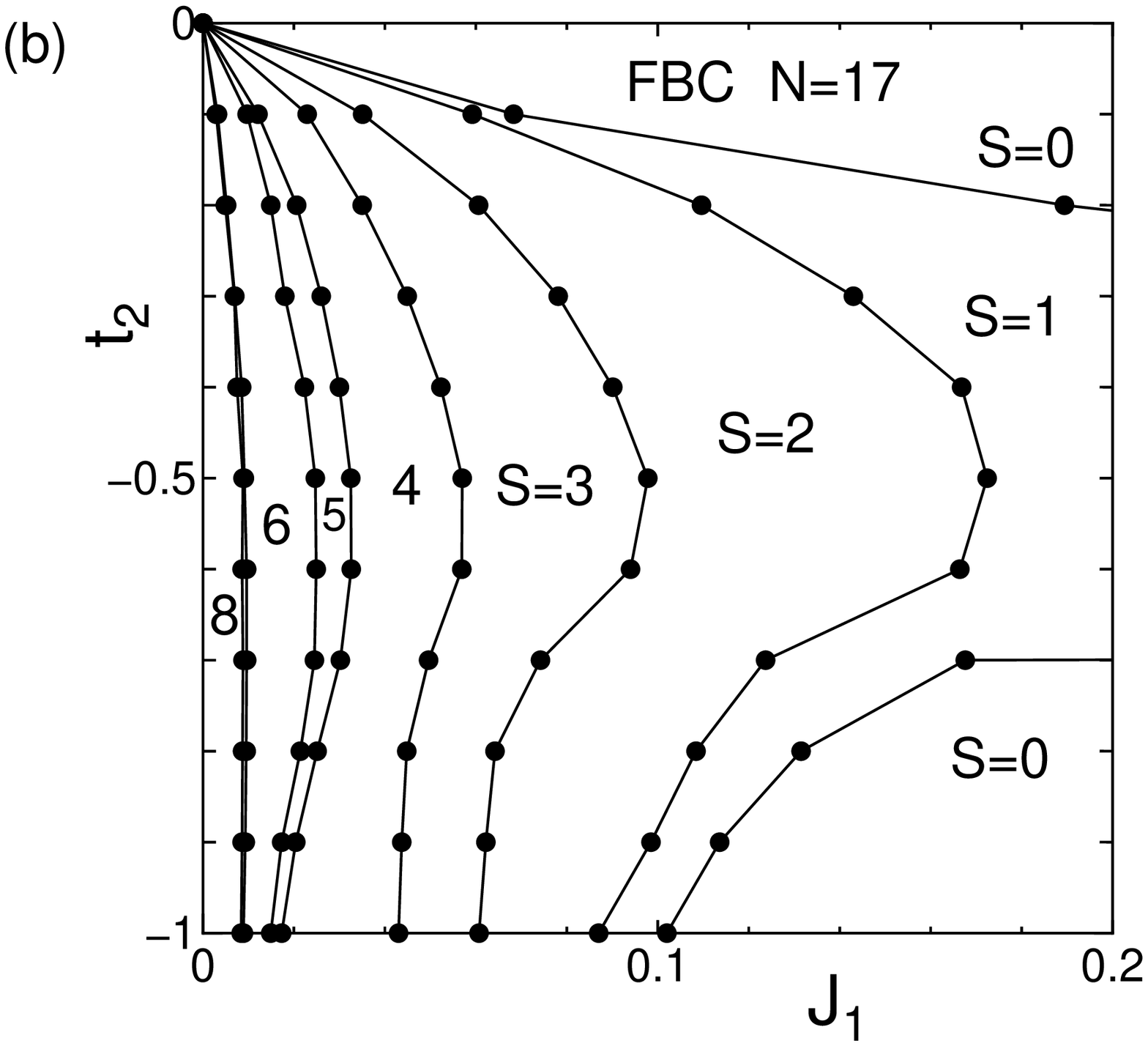}
\includegraphics[width=0.7 \linewidth]{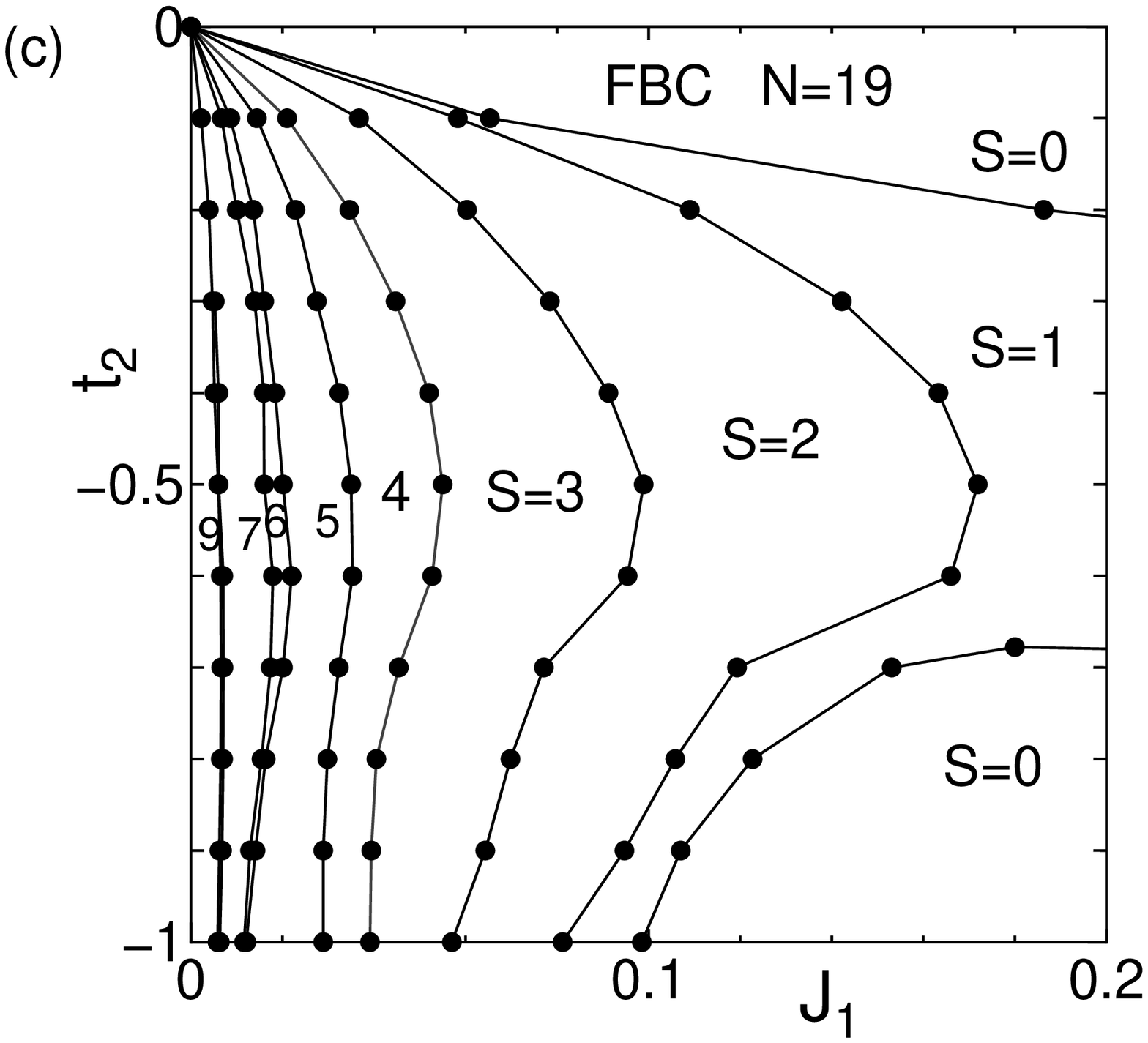}
\end{center}
\caption{Ground-state phase diagrams for the one-hole case in systems with (a) 15, (b) 17, and (c) 19 sites under the FBC.}
\label{phase_fbc}
\end{figure}

We numerically diagonalized the Hamiltonian (\ref{tJ-Ham}) in the one-hole case with up to 23 sites using the standard Lanczos algorithm, and obtained the ground states and low excited states. 
The ground-state phase diagrams for the systems of $N=$15, 17, and 19 under the free boundary condition (FBC) are shown in Fig. \ref{phase_fbc}. 
The scale of the $J_1$ axis is expanded relative to that of the $t_2$ axis to exhibit a narrow regime of $J_1 \stackrel{<}{_\sim} 0.2$. 

%<< FIG.1
%>>

%<< FIG.2
\begin{figure}[t]
\begin{center}
\includegraphics[width=0.7 \linewidth]{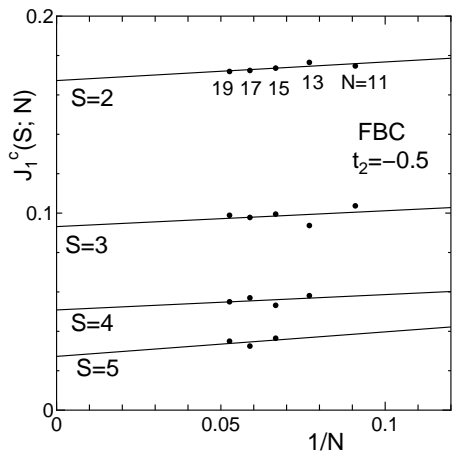}
\end{center}
\caption{System-size dependences $J_1^c(S; N)$ of the phase boundaries between the phases of $S =$ 2, 3, 4, and 5 under the FBC at $t_2 = -0.5$.}
\label{size_phase_bound}
\end{figure}
%>>

Each phase diagram in Fig. \ref{phase_fbc} includes a series of partial ferromagnetic phases with $S =$ 2, 3, 4, $\cdots, S_\mathrm{max}$, where $S_\mathrm{max}$ = $(N-1)/2$. 
The phase shapes seem to be almost independent of the system-size $N$. 
To ensure that the partial ferromagnetic phases survive in the large system-size limit, we extrapolate boundary values of $J_1$ between adjacent partial ferromagnetic phases. 
We denote the boundary value of $J_1$ between the phases of $S-1$ and $S$ at $t_2 = -0.5$ for system size $N$ as $J_1^c(S; N)$ and their large system-size limit by $J_1^c(S)$. 
The values of $J_1^c(S; N)$ for $S=$ 2, 3, 4, and 5 are plotted against $1/N$ in Fig.~\ref{size_phase_bound}. 
The extrapolation is performed by assuming the system-size dependence to be 
%-------------------------------------------------------------
\begin{align}  %Eq_(4)
J_1^c(S; N) = J_1^c(S)+ \frac{A(S)}{N} , 
\label{extrapolation}
\end{align} 
%-------------------------------------------------------------
where $A(S)$ is a fitting parameter for each $S$. 
Using this fitting method, we evaluated $J_1^c(S)$ as 0.167, 0.093, 0.051, and 0.027 for $S=$ 2,  3, 4, and 5, respectively. 
The corresponding values of $A(S)$ are 0.0804, 0.0945, 0.0779, and 0.1244 for $S=$ 2, 3, 4, and 5, respectively. 
In Fig.~\ref{size_phase_bound}, Eq.~(\ref{extrapolation}) is also shown by a solid line for each $S$. 
The nonzero values of $J_1^c(S)$ numerically confirm that the partial ferromagnetic phases of $S=$ 2, 3, 4, and 5 survive in the large system-size limit. 

We also calculated the value of $J_1^c(1; N)$ for each $N$. 
This is the value of $J_1$ between the singlet phase for large $J_1$ and the phase of $S=1$ 
at $t_2 = -0.5$. 
The same equation for extrapolation, Eq.~(\ref{extrapolation}), gives the extrapolated value $J_1^c(1) = 0.84$ with $A(1) = -0.6503$. 

%<< FIG.3
\begin{figure}[t]
\begin{center}
\includegraphics[width=0.7 \linewidth]{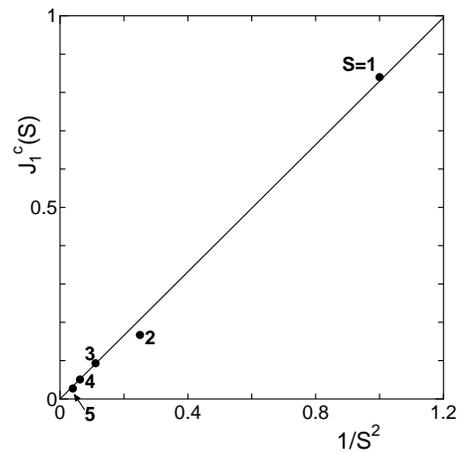}
\end{center}
\caption{$S$-dependence of the boundary value, $J_1^c(S)$, between the phases of total spins $S$ and $S-1$ at $t_2 = -0.5$.}
\label{Jc_S}
\end{figure}
%>>

The $S$-dependence of $J_1^c(S)$ is shown in Fig.~\ref{Jc_S} and is roughly described by the equation 
%-------------------------------------------------------------
\begin{align}  %Eq_(5)
J_1^c(S) \simeq \frac{0.8}{S^2} . 
\label{J1cS}
\end{align} 
%-------------------------------------------------------------
This equation shows that the complete ferromagnetic ground state with $S=\Smax$ appears only at $J_1=0$ in the large system-size limit; then, we also have  $J_2=0$ owing to the choice of parameter restriction. 
When $J_1$ continuously changes from a large value to zero, the system in a singlet ground state undergoes successive quantum phase transitions with increasing total spin $S$ in steps of one, and finally reaches a complete ferromagnetic state at $J_1 = 0$. 
In the case of $J_1 = 0$, electrons in the system interact with each other only through infinite on-site repulsions. 
The present model then becomes a 1D Hubbard model with first and second transfers and infinite on-site repulsions. 
This complete ferromagnetism is then exactly the Nagaoka ferromagnetism.\cite{Nagaoka,Thouless,Daul,White,Eisenberg,Park} 

Until now, we have performed numerical diagonalization using the FBC. 
The FBC is advantageous for the purpose of extrapolation, because many physical quantities change smoothly with system size $N$ under the FBC. 
However, the FBC is not adequate for calculating correlation functions because of the lack of translational invariance. 
Hence, in the next section, we adopt the anti-periodic boundary condition (APBC)\cite{APBC} to numerically calculate correlation functions for odd $N$. 

%<< FIG.4
\begin{figure}[t]
\begin{center}
\includegraphics[width=0.68 \linewidth]{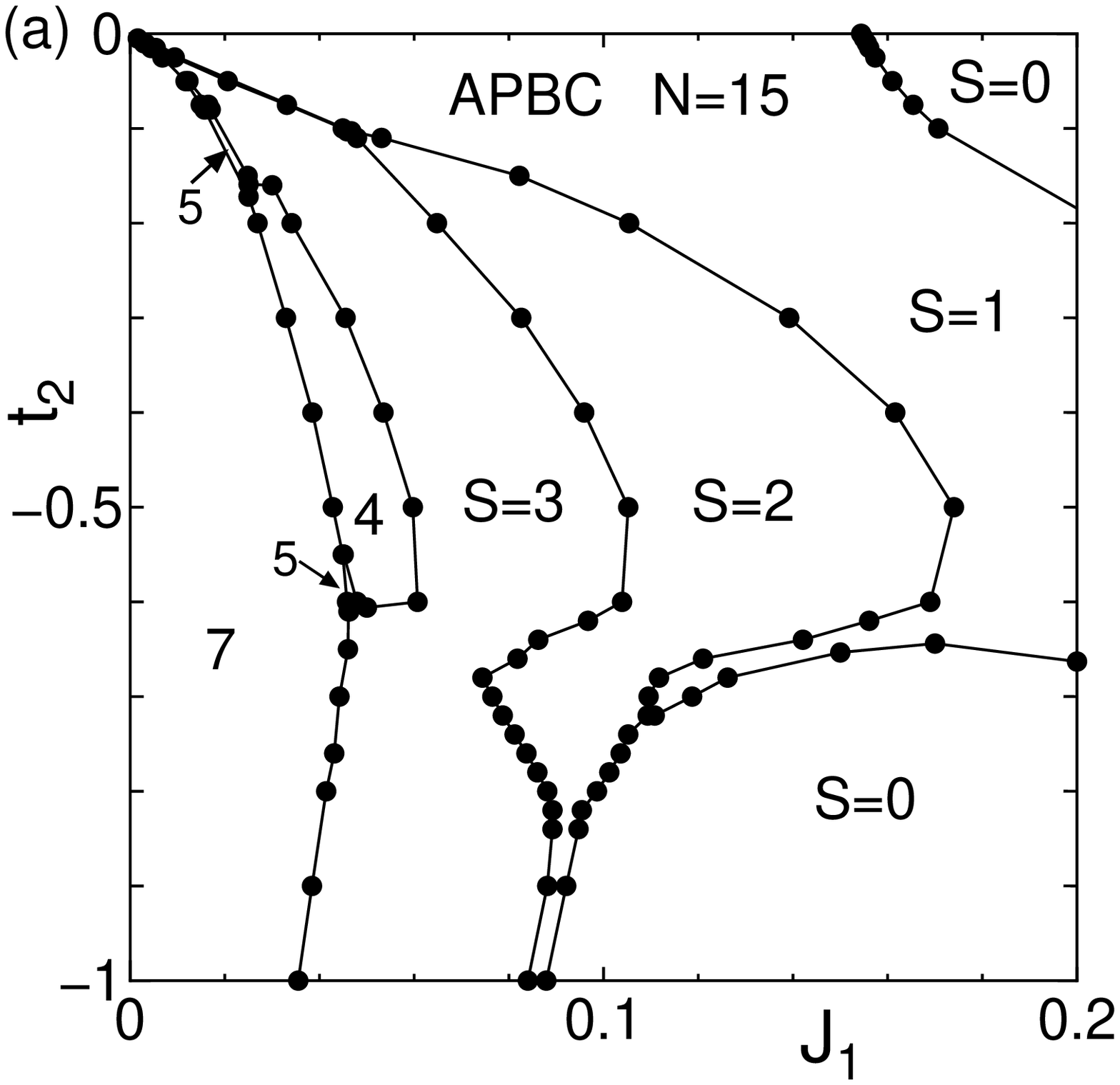}
\includegraphics[width=0.68 \linewidth]{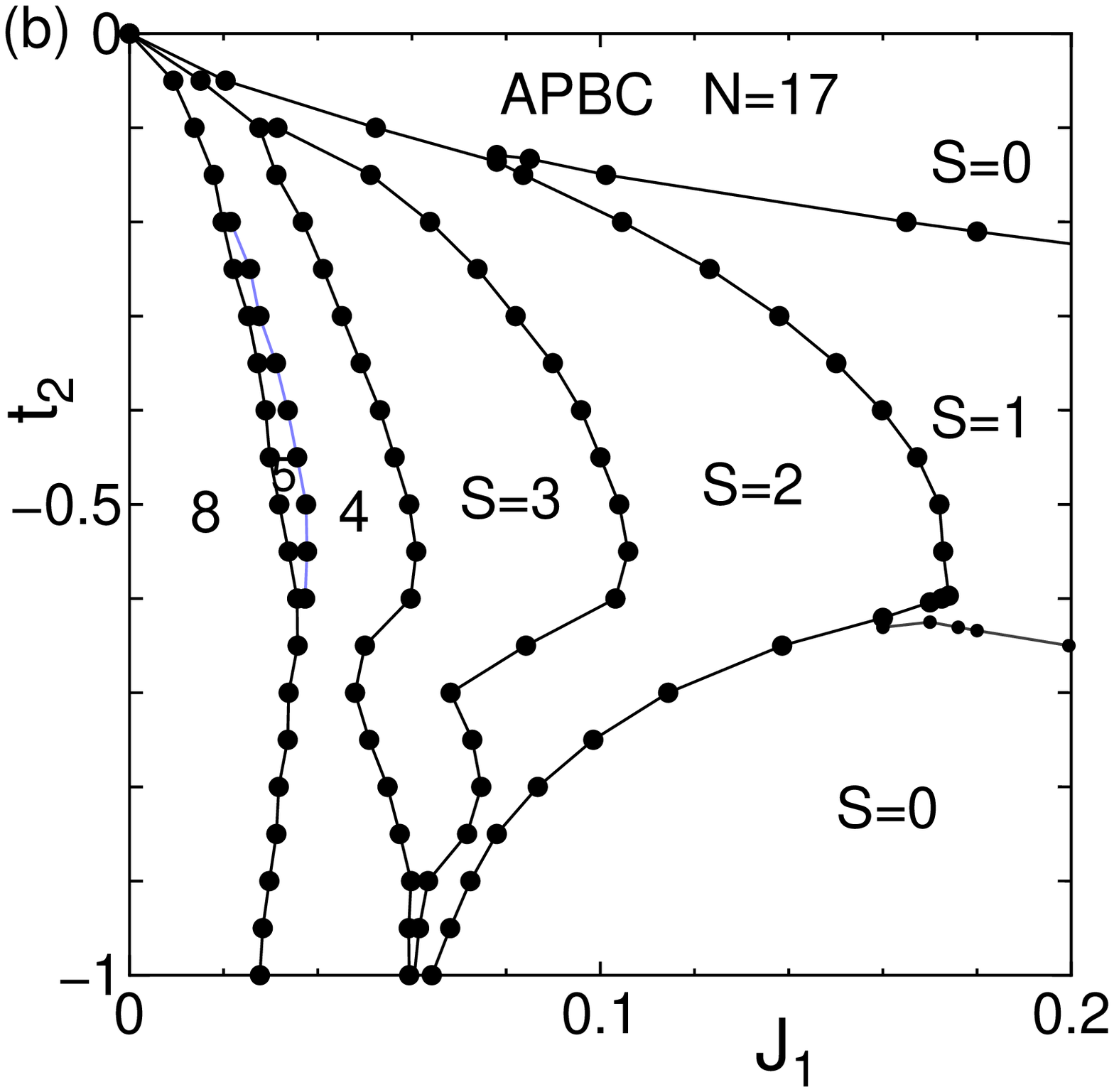}
\includegraphics[width=0.68 \linewidth]{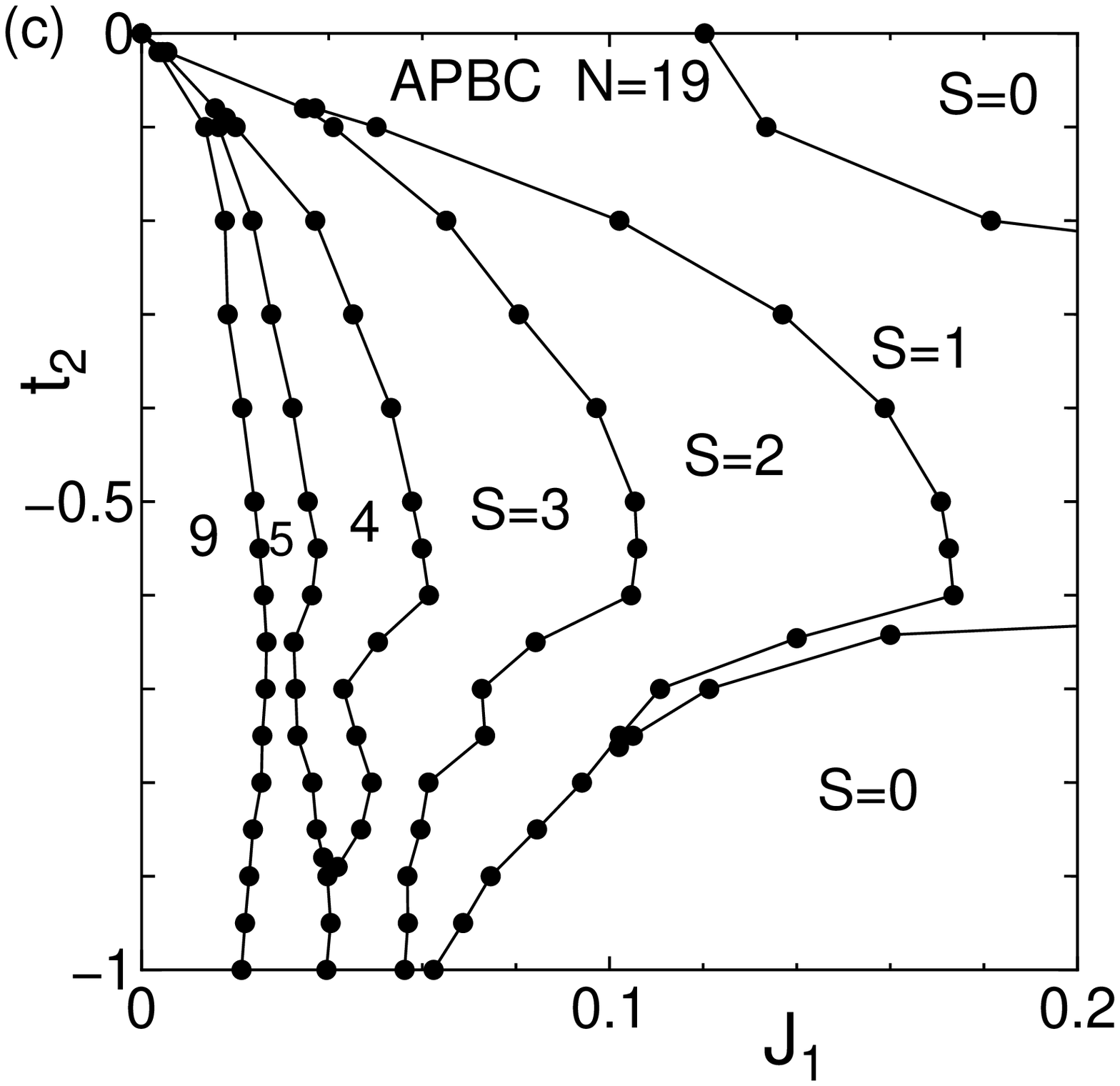}
\end{center}
\caption{Ground-state phase diagrams for the one-hole case in systems with (a) 15, (b) 17, and (c) 19 sites under the APBC.}
\label{phase_apbc}
\end{figure}
%>>

To ensure use of a parameter regime in which the APBC works correctly, 
we calculated the ground-state phase diagrams for the systems with $N=$ 15, 17, and 19 under the APBC. 
The results are shown in Fig. \ref{phase_apbc}. 
Comparing them to those under the FBC in Fig. \ref{phase_fbc}, the boundaries between partial ferromagnetic phases are similar for $-0.6 \lesssim t_2 \lesssim 0$. 
In this regime, the phase boundaries are not supposed to be seriously affected by the choice of boundary condition, even if $N$ is as small as 15. 
We will perform the numerical diagonalization to calculate a correlation function and other quantities under the APBC at $t_2 = -0.5$ as a typical case. 

\bigskip\bigskip

%%%%%%%%%%%%%%%%%%%%%%%%
\section{Ferromagnetic clouds in the partial ferromagnetic ground states}\label{msbstate}

We examine the correlation between the charge and spin degrees of freedom in partial ferromagnetic phases. 
The charge degree of freedom is represented by the hole density operator defined as 
%-------------------------------------------------------------
\begin{align} %Eq_(6)
\nhi &= 1- n_{i,\uparrow} - n_{i,\downarrow} 
\label{nhi_def}
\end{align} 
%-------------------------------------------------------------
with $n_{i,\uparrow} = c_{i,\uparrow}^{\dagger}c_{i,\uparrow}$ and $n_{i,\downarrow} = c_{i,\downarrow}^{\dagger}c_{i,\downarrow}$. 
The spin degree of freedom is represented by the spin density operator of the $z$-component defined as %-------------------------------------------------------------
\begin{align} %Eq_(7)
s_z(i) &= \frac{1}{2} (n_{i,\uparrow}-n_{i,\downarrow}) . 
\label{szi_def}
\end{align} 
%-------------------------------------------------------------
Then, the correlation function between the hole and a spin is defined as 
%-------------------------------------------------------------
\begin{align} %Eq_(8)
\Chsr = \langle \nhi \, s_z(i+r) \rangle, 
\label{chsr_def}
\end{align} 
%-------------------------------------------------------------
which is calculated under the APBC. 
It represents the spin density at the relative distance $r$ from the hole position $i$. Note that   $\Chsr$ is independent of the site $i$ because of the translational invariance. 
We calculate $\Chsr$ for partial ferromagnetic phases with $S_z = S$ without destroying generality, with the total spin directed along the $z$-axis. 

%<< FIG.5
\begin{figure}[b]
\begin{center}
\includegraphics[width=0.7 \linewidth]{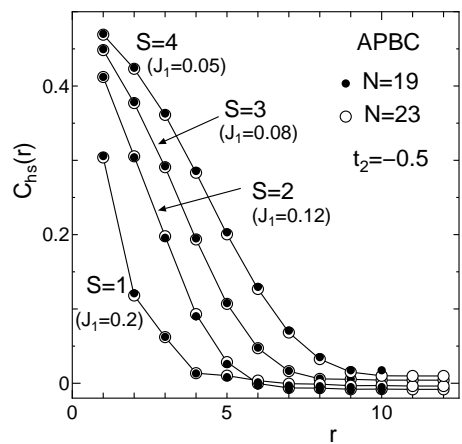}
\end{center}
\caption{The correlation functions $\Chsr$ of the hole and a spin as functions of the distance between them. 
The plotted data are in the phases of $S =$ 1, 2, 3, and 4 for the system with $N=19$ and $23$ under the APBC.}
\label{correlation_hs}
\end{figure}
%>>

In Fig. \ref{correlation_hs}, we show the correlation function $\Chsr$ of partial ferromagnetic states in the phases of $S =$ 2, 3, and 4 for the systems with $N=$ 19 and 23. 
The points where we calculate the correlation function are $(J_1, t_2)$ = 
$(0.12, -0.5)$ for $S=2$, 
$(0.08, -0.5)$ for $S=3$, and 
$(0.05, -0.5)$ for $S=4$. 
We have added the result at $(J_1, t_2)$ = $(0.2, -0.5)$ in the phase of $S=1$ for comparison. 
Figure \ref{correlation_hs} shows that the data for $N =$ 23 almost agrees with the data for $N=$ 19, meaning that the finite size effect is very small for the system with $N =$ 19. 
The result for $\Chsr$ shows that the spin density decreases rapidly with distance $r$, as measured from the hole position in each of the partial ferromagnetic states of $S =$ 2, 3, and 4, and the rapidity is weaker in these states than in the $S=1$ state. 
This means that polarized electronic spins are drawn around the hole in a partial ferromagnetic state. 

We quantitatively analyze the $\Chsr$ by assuming a Gaussian-type decay. 
The fitting function $\tChsr$ is given as 
\begin{align}
\tChsr = A e^{-r^2/\lambda^2} 
\label{corr_fit}
\end{align}
with fitting parameters $A$ and $\lambda$. 
The parameter $\lambda$ represents the size of the assembly of polarized spins drawn around the hole and is also the moving range of the hole in the ferromagnetic regime. 
By fitting, the values of $(A, \lambda)$ were found to be 
$(0.295, 0.20)$ for $S=1$, 
$(0.45, 0.096)$ for $S=2$, 
$(0.48, 0.059)$ for $S=3$, and 
$(0.49, 0.036)$ for $S=4$ 
at the same points of $(J_1, t_2)$ as those used for calculating $\Chsr$ in the last section. 
These results indicate that spins in the same direction collect around the hole to form an assembly of spins. 

%<< FIG.6
\begin{figure}[t] 
\begin{center}
\includegraphics[width=0.6 \linewidth]{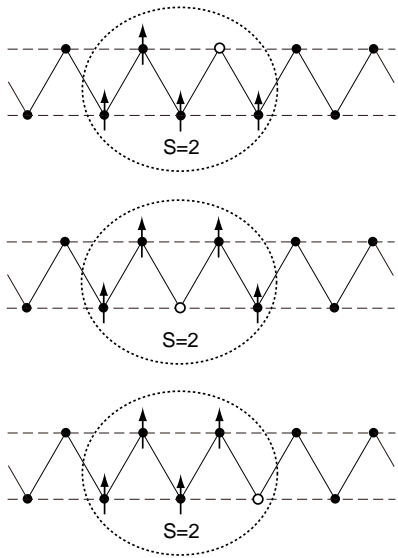}
\end{center}
\caption{
Typical configurations of the hole and spins included in the ground state of the $S=2$ phase. 
The section enclosed by a dotted loop is a bound state of the hole and four up spins forming an $S=2$ ferromagnetic cloud. 
The background electronic spins outside form a singlet spin liquid. 
The ferromagnetic cloud moves as a composite particle, and hence the wave function also includes patterns of the bound state at different locations.}
\label{typical_config}
\end{figure}
%>>

Figure \ref{typical_config} shows pictures of the typical configuration patterns of the hole and spins included, as parts of a linear combination, in the $S=2$ partial ferromagnetic state. 
Inside the dotted loop, the electronic spins form a ferromagnetic state with the hole.
The other electronic spins outside the ferromagnetic state form a background singlet state in a partial ferromagnetic state.
We call this spin assembly in the partial ferromagnetic state  the {\it ferromagnetic cloud}.

The above analysis  based on the correlation function shows
that the hole is confined within the ferromagnetic cloud. 
In other words, the spin degrees of freedom of electrons are bounded by a force mediated by the hole.
In the present case, the ferromagnetic mechanism only works locally near the hole. 
We consider that the mechanism of local ferromagnetism is basically the same as that of Nagaoka ferromagnetism.

%<< FIG.7
\begin{figure}[t] 
\begin{center}
\includegraphics[width=0.7 \linewidth]{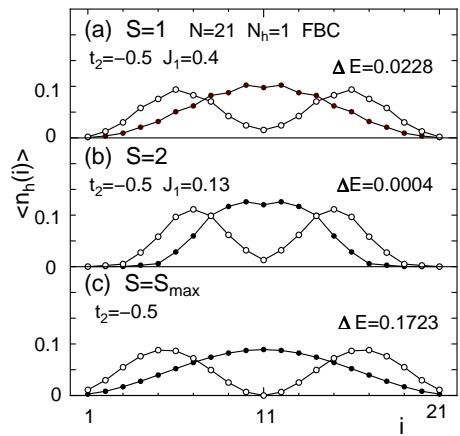}
\end{center}
\caption{Hole density $\langle \nhi \rangle$ as a function of site $i$ under the FBC (a) for $S=1$ at $t_2=-0.5$ and $J_1=0.4$, (b) for $S=2$ at $t_2=-0.5$ and $J_1=0.13$, and (c) for $S=\Smax$ at $t_2=0$. 
Here, the solid circles are data for the ground state and the open circles are for the excited states. }
\label{hole_density}
\end{figure}
%>>

We analyze the translation of a ferromagnetic cloud as a whole by using the analogy of a single particle. 
We take the hole position $i$ as a representative position of the ferromagnetic cloud. 
Then the relative motion $r$ between the center of the spins and the hole is interpreted as inner motion within the ferromagnetic cloud. 
The spatial distribution of the hole position appears in the hole density $\langle \nhi \rangle$. 
We calculated the hole density in the $S=2$ partial ferromagnetic state and show the results in panel (b) of Fig.~\ref{hole_density} for the $N=21$ system under the FBC. 
For comparison, we also show the densities of $S=1$ and $S=\Smax$ in panels (a) and (c), respectively, of the same figure. 
The hole densities are for the ground and first excited states in each panel. 
The parameters are set as $(J_1, t_2)$ = $(0.4, -0.5)$ for the $S=1$ state, $(0.13, -0.5)$ for the $S=2$, and $(0, 0)$ for $S=\Smax$. 
We see commonly in each panel, that the hole density in the ground state has no node and the first excited state has a single node. 

The FBC is expected to work for a composite particle in the same manner as a square-well potential works for a single particle in elementary quantum mechanics. 
If a magnetic cloud moves with an effective mass $M_S$ freely in the potential well, the energy is $E_m(S) = \hbar^2 k_m^2/(2M_S)$, where $k_m = m\pi/N$ is the wave number for quantum number $m$. 
Then the square of the wave function behaves as $\rho_m(i) \propto \sin^2(k_m i)$. 
In fact, the panel (b) of Fig.~\ref{hole_density} shows that the densities of the ground and first excited states are similar to $\rho_1(i)$ and $\rho_2(i)$, respectively, with the same numbers of nodes. 
The states of $S=\Smax$ in the panel (c) of Fig.~\ref{hole_density} are references, since a hole in the 1D ferromagnetic state is exactly a single quantum-mechanical particle. 

The effective masses of the composite particle of $S=1$ and ferromagnetic clouds with $S \ge 2$ are represented by numerically calculated energies. 
Using the free particle formula for the energies $E_m(S)$ of the ground and first excited states, the relative effective mass $M_S^* \equiv M_S/M_{\Smax}$ is given as 
\begin{align}
M_S^* = \frac{E_2(\Smax) - E_1(\Smax)}{E_2(S) - E_1(S)} . 
\label{mass_eff}
\end{align}
We substitute numerically obtained values of the energies into Eq.~(\ref{mass_eff}) and estimated the masses of the composite particle and the ferromagnetic cloud with $S=2$ as $M_1^* \simeq 7.5$ and $M_2^* \simeq 410$, respectively. 
The ferromagnetic cloud with $S=2$ is extremely heavy and moves very slowly in comparison with a single hole in the ferromagnetic background with $S = \Smax$. 

%\bigskip

%%%%%%%%%%%%%%%%%%%%%%%
\section{Summary and Discussion}\label{disc}

We numerically investigated partial ferromagnetic phases with total spins $S\ge 2$ in the one-hole case of the 1D $t_1$-$t_2$-$J_1$-$J_2$ model. 
The partial ferromagnetic phases exist in the large system-size limit and are stacked in a regime of positive small $J_1$ in the $J_1$-$t_2$ plane. 
We find that the ground state in each partial ferromagnetic phase includes a ferromagnetic cloud, which is a multiple-spin bound state together with the hole. 
The ferromagnetic cloud is a large magnetic polaron with a heavy mass in a single-band electronic system. 
It is formed owing to hole motion and is explained as local Nagaoka ferromagnetism. 

One question that remains is whether or not similar magnetic polarons are formed in higher dimensions. 
We have found partial ferromagnetic phases for small systems in the two-dimensional $t_1$-$t_2$-$J_1$-$J_2$ model,\cite{Sano2011} although we have not yet confirmed the existence of magnetic polarons in the large system-size limit.  
However, we think that partial ferromagnetic phases with magnetic polarons are formed in two- and three-dimensional systems, 
since Nagaoka ferromagnetism does not seem to depend on the lattice dimensionality.\cite{Nagaoka,Thouless,Daul,White,Eisenberg,Park} 
Several authors have briefly discussed the existence of magnetic polarons in two- or three-dimensional systems in the context of Nagaoka ferromagnetism.\cite{NF-polaron} 
However, they have not provided clear results, such as concrete conditions for forming a magnetic polaron and the properties of such a polaron. 
We think that further studies are required to answer these problems. 

Finally, we consider the thermodynamic limit in which an infinite system includes a finite and low density of holes. 
In an appropriate parameter regime, it is possible that the holes form almost independent ferromagnetic clouds. 
As for two $S$=1 composite particles, we have numerically shown that these particles coexist and repel each other at short range.\cite{Sano2011} 
For two ferromagnetic clouds with $S \ge 2$, we could not carry out similar numerical analysis since their sizes are much larger than the $S$=1 composite particles. 
A remaining problem is whether ferromagnetic clouds remain independent objects or coalesce to form a large ferromagnetic area including holes. The latter scenario implies a phase separation. 

If ferromagnetic clouds maintain themselves even at a finite low density, 
it seems that they form a ferromagnetic colloidal solution as known as magnetic fluid or ferrofluid.\cite{Sano1983,Raj} 
Since the ferromagnetic clouds are quantum mechanical, 
an assembly of them is a quantum ferrofluid. 
There is then a question of whether the magnetic clouds are bosonic, fermionic, or obey other statistics. 
It has been reported that ultracold atoms with large spins form a quantum ferrofluid in experimental investigations.\cite{Lahay} 
Theoretically, a similar state has been found in a Hubbard model with up to second transfer.\cite{Nishimoto}

\section*{ACKNOWLEDGMENTS}

This work was supported by JPSJ KAKENHI Grant Number 26400411.


\begin{references}
\bibitem{mpolaron}
The present magnetic polaron is different from a magnetic polaron formed by the exchange interaction between the spin of a carrier electron and the localized spin of a magnetic ion in, e.g., the $s$-$d$ model.
\bibitem{APBC} 
The APBC brings a closed shell to spin polarized electrons in the non-interacting band. 
It stabilizes the complete ferromagnetic state in the limit of $J_1 \to 0$, 
while normal periodic boundary conditions destabilize the complete ferromagnetic state at sufficiently small $J_1$.
\end{references}
\end{document}